\documentstyle[12pt,epsfig]{article}
\newlength{\dinwidth}
\newlength{\dinmargin}
\newcommand{\ba}{\begin{array}}
\newcommand{\ea}{\end{array}}
\newcommand{\beq}{\begin{equation}}
\newcommand{\eeq}{\end{equation}}
\newcommand{\bea}{\begin{eqnarray}}
\newcommand{\eea}{\end{eqnarray}}

\def\bce{\begin{center}}
\def\ece{\end{center}}

\def\nonu{\nonumber}

\def\pa{\partial}

\def\de{\delta}

\def\ep{\epsilon}

\def\La{\Lambda}

\def\eps6{{\displaystyle \mathop{\epsilon}^{6}}{}}

\def\nab6{{\displaystyle \mathop{\nabla}^{6}}{}}

\begin{document}
\thispagestyle{empty}
\addtocounter{page}{-1}
\begin{flushright}
{\tt hep-th/0212231}\\
\end{flushright}
\vspace*{1.3cm}
\centerline{\Large \bf Supersymmetric $SO(N_c)$ Gauge Theory and
Matrix Model}
\vspace*{1.5cm}
\centerline{{\bf Changhyun Ahn}$^1$ and {\bf Soonkeon Nam}$^2$}
\vspace*{1.0cm}
\centerline{\it $^1$Department of Physics,
Kyungpook National University, Taegu 702-701, Korea}
\vspace*{0.2cm}
\centerline{\it $^2$Department of Physics and Research Institute for Basic
Sciences,}
\centerline{ \it Kyung Hee University, Seoul 130-701, Korea}
\vspace*{0.8cm}
\centerline{\tt ahn@knu.ac.kr, \qquad nam@khu.ac.kr}
\vskip2cm
\centerline{\bf Abstract}
\vspace*{0.5cm}

By applying the method of Dijkgraaf-Vafa,
we study matrix model related to supersymmetric $SO(N_c)$ gauge theory
with $N_f$ flavors of quarks in the vector representation found by
Intriligator-Seiberg.
By performing the matrix integral over tree level superpotential
characterized by light meson fields (mass deformation)
in electric theory, we reproduce the
exact effective superpotential in the gauge theory side. Moreover,
we do similar analysis in magnetic theory.
It turns out the matrix descriptions of both electric and magnetic theories
are the same: Seiberg duality in the gauge theory side.

\vspace*{\fill}


\baselineskip=18pt
\newpage
\renewcommand{\theequation}{\arabic{section}\mbox{.}\arabic{equation}}

\section{Introduction}
\setcounter{equation}{0}

Recently, Dijkgraaf and Vafa \cite{dv}
have proposed a technique for calculating
the effective superpotential for the glueball superfield in an ${\cal N}=1$
gauge theory through the planar diagrams of matrix model.
Several tests have been performed for
this proposal in \cite{cm}-\cite{seiberg}.
In particular, the theories
with matter fields in the fundamental representation have been found in
\cite{acfh1,mcg,suz,br1,dj,tachik,feng,fh,ohta,bhr,hof,suz1,seiberg}.

In this letter, we compute the matrix integral over tree level superpotential
characterized by light meson fields (mass deformation)
in electric ${\cal N}=1$ $SO(N_c)$ gauge theory with
$N_f$ flavors of quarks in the vector (fundamental) representation found by
Intriligator-Seiberg \cite{is} and  we reproduce the
exact effective superpotential in the gauge theory side. Moreover,
we do similar analysis in magnetic theory.
The matrix descriptions of both electric and magnetic theories
are coincident with each other implying Seiberg duality in the gauge
theory side. In \cite{fo,ino,ookouchi,ashoketal,feng1,jo},
there are some relevant
works on the gauge group $SO(N_c)$ in the view point of matrix
model. For $U(N_c)$ gauge theory with $N_f$ flavors of quarks in the
fundamental and anti-fundamental representations,
the matrix descriptions of both electric
and magnetic theories have been found in \cite{feng,fh}.

\section{Matrix model description of electric theory }
\setcounter{equation}{0}

Let us
 deform our ${\cal N}=1$ supersymmetric $SO(N_c)$ gauge theory
with $N_f$ flavors of quarks $Q^j_a(j=1, 2, \cdots, N_f, a=1, 2, \cdots,
N_c)$ in the vector (fundamental)
representation( $N_c \geq 4, N_f \leq N_c-5$) \cite{is},
by adding the mass terms  of gauge invariant meson superfields
$M^{ij}=Q^i \cdot Q^j$
\bea
W_{\mbox{tree}} = \frac{1}{2} \mbox{Tr} \; m M =
\frac{1}{2} \sum_{j=1}^{N_f} m_j \sum_{a=1}^{N_c} Q^j_a Q^j_a
\label{tree}
\eea
where
it is understood that
quark superfields are represented
in an $N_f \times N_c$ matrix form in terms of the flavor
and color indices. In the gauge theory side, one can make this superfield
matrix diagonal by using the gauge and global rotation.

The idea is to use this tree level superpotential as the potential
for the matrix model. At first, let us consider $SO(N)$ matrix
model at large $N$ by replacing the gauge theory fields with $N_f
\times N$ matrices in order to calculate the contributions to the
free energy. Then the partition function can be written as \bea Z=
\frac{1}{\mbox{Vol}(SO(N))} \left(\frac{\La} {2\pi
g_s}\right)^{\frac{1}{2}N_f N} \int \prod_{j=1}^{N_f} [d Q^j_a]
e^{-\frac{1}{2g_s} \sum_{j=1}^{N_f} m_j \sum_{a=1}^N Q^j_a Q^j_a}
\nonu \eea where we put the  factor $\left(\frac{\La} {2\pi
g_s}\right)^{\frac{1}{2}N_f N}$ in order to make the integration
measure dimensionless\footnote{In fact we can put any dimensionful
term with the right dimension  inside the parenthesis, but this
particular form gives the right form of the superpotential for the
gauge theory.} and $\La$ is the scale of $SO(N_c)$ gauge theory
and $m_j$ is the mass of $j$-th quark. In this gauge theory there
is no difference between left- and right-handed quarks which is
different from $SU(N_c)$ gauge theory where the right-handed quark
is the left-handed anti-quark. Recall that the superpotential
should have $R$-charge 2 and dimension 3.

The large $N$-limit behavior of the volume for $SO(N)$ group,
the normalization of the matrix path integral,
 can be read off from  \cite{mac,ov,ashoketal}
\bea
\log \mbox{Vol}(SO(N)) & = &
\log \frac{\sqrt{2} \left(2\pi\right)^{N^2/4}}{\left
(N-3\right)!\left(N-5\right)! \cdots 3!1!\left(N/2-1\right)!}
\nonu \\
& = & -\frac{1}{4} N^2 \log N +
\frac{1}{4} N^2 \left( \frac{3}{2} + \log \pi +\log 2 \right)+
\frac{N}{4} \log N \nonu \\
&  & +\frac{1}{24} \log N +\frac{N}{4} \left( -1+\log 2-\log \pi \right)
+ {\cal O}(1).
\nonu
\eea
The usual matrix integral for flavor part is
\bea
 \int \prod_{j=1}^{N_f} [d Q^j_a]
e^{-\frac{1}{2g_s} \sum_{j=1}^{N_f} m_j \sum_{a=1}^N Q^j_a Q^j_a} =
\left( \frac{\pi^{N_f} g_s^{N_f}}{ \mbox{det} m} \right)^{N/2}.
\nonu
\eea

In the large $N$-limit we are interested in, the glueball
superfield $S$ can be identified with $g_s N$  as a second step
and one can write the log of the partition function as follows:
\bea \log Z &=& \frac{1}{4} N^2 \log \left(\frac{N}{2\pi
e^{3/2}}\right) + \frac{N}{2} N_f \log \La -\frac{N}{2}
 \log \mbox{det} m \nonu \\
& = &  \frac{S^2}{4g_s^2} \log \left( \frac{S}{2\pi e^{3/2} g_s}
\right) + \frac{S}{2g_s} N_f \log \La - \frac{S}{2g_s}
 \log \mbox{det} m
\nonu \\
& \equiv & -\frac{1}{g_s^2} {\cal F}_2 -\frac{1}{g_s} {\cal F}_1.
\nonu \eea Note that all the effects of the matter is in ${\cal
F}_1$.

Then
the effective superpotential for the glueball field $S$ can be
computed as the derivative of the contribution to free energy plus
the contribution from flavors
 \cite{dv,acfh1}
\bea
W  & = &
\left(N_c-2\right) \frac{\pa {\cal F}_2}{\pa S} + {\cal F}_1
\nonu \\
& = & \frac{1}{2} \left(N_c-2 \right) \left( S - S \log \left(
\frac{S}{ \La^3} \right)   - \frac{S}{N_c-2} N_f \log \La +
\frac{S}{N_c-2} \log \mbox{det} m \right)
\nonu \\
& = & \frac{1}{2} \left(N_c-2 \right) \left( S -S \log
\frac{S}{\left( \La^{3(N_c-2)-N_f} \mbox{det}
m\right)^{1/\left(N_c-2\right)} }  \right). \label{W} \eea Here we
have used the results that the derivative of ${\cal F}_2$ gives
the Veneziano-Yankielowicz superpotential\cite{dv}.

Solving the F-flatness condition $\pa_S W=0$ (minimizing $W$ with
respect to a glueball superfield $S$) one gets, by the scale
normalization $\La^{3\left(N_c-2\right)-N_f} = 16
\La^{3\left(N_c-2\right)-N_f}_{N_c, N_f} $, \bea S = \left(16
\La^{3\left(N_c-2\right)-N_f}_{N_c,N_f} \mbox{det} m
\right)^{1/\left(N_c-2\right)} \ep_{N_c-2}, \;\; \qquad
\ep_{N_c-2} = e^{2\pi i k/\left(N_c-2\right)},  \;\; k=1, \cdots,
\left(N_c-2\right) \nonu \eea with the phase factor $e^{2\pi i
k/\left(N_c-2\right)}, k=1, \cdots, \left(N_c-2\right)$ reflecting
the $(N_c-2)$ supersymmetric vacua but physically equivalent vacua
of the theory coming from the spontaneous breaking of a discrete
symmetry. Then the exact superpotential by plugging $S$ into
(\ref{W})( integrating out the $S$) leads to \bea W & = &
\frac{1}{2} \left(N_c  -2 \right) \ep_{N_c-2} \left( 16
\La_{N_c,N_f}^{3\left(N_c-2\right)-N_f} \mbox{det}  m
\right)^{1/\left(N_c-2\right)}
\nonu \\
&= &   \frac{1}{2} \left(N_c  -2 \right) \ep_{N_c-2}
\hat{\La}_{N_c-N_f,0}^3
\label{eW}
\eea
where
$\hat{\La}_{N_c-N_f,0}^3=  \left(
16 \La_{N_c,N_f}^{3\left(N_c-2\right)-N_f} \mbox{det}  m
\right)^{1/\left(N_c-2\right)}$ is the low
energy scale \cite{is} for the $SO(N_c-N_f)$ Yang-Mills theory.
This is nothing but ADS superpotential below for pure $SO(N_c)$
Yang-Mills gauge theory.
In the low energy effective theory, the classical vacuum degeneracy was
lifted by quantum effects which is represented by a dynamically
generated superpotential for the light meson fields $M^{ij}$.
The superpotential \cite{is} generated by gaugino condensation leads to
\bea
W_{ADS} =
\frac{1}{2} \left(N_c -N_f -2 \right) \ep_{N_c-N_f-2} \left(
\frac{16 \La_{N_c,N_f}^{3\left(N_c-2\right)-N_f}}{\mbox{det}
M }\right)^{1/\left(N_c-N_f-2\right)}
\nonu
\eea
where
$ \ep_{N_c-N_f-2}$ is the $(N_c-N_f-2)$-th root of unity.
This theory has no vacuum but by adding the mass terms (\ref{tree})
to the superpotential $W_{ADS}$,
the theory has $(N_c-2)$ supersymmetric vacua.
If not all of the matter fields are massive, in the gauge theory side,
one can integrate  out massive quarks and get the effective superpotential
at low energy for the massless ones.
It is the same form as the above $W_{ADS}$ but with the scale replaced by
the low energy one.

\section{Matrix model description of magnetic theory }
\setcounter{equation}{0}

In the IR theory of electric theory in previous section, the magnetic theory
is described by an $SO(\widetilde{N}_c=N_f-N_c+4)$ gauge theory
( $N_f > N_c, N_c \geq 4 $ )
with $N_f$ flavors of dual quarks $q^j_a(j=1, \cdots, N_f, a=1, \cdots,
\widetilde{N}_c)$ and the additional
gauge singlet fields $M^{ij}$ which is an elementary field of dimension 1
at the UV-fixed point \cite{is}.
The matter field variables in the magnetic theory
are the original electric variables $M^{ij}$ and magnetic quarks
$q_i$ with the superpotential together with mass term
\bea
W = \frac{1}{2 \mu} M^{ij} q_i \cdot q_j + \frac{1}{2}  \mbox{Tr} \; m M.
\nonu
\eea

After integrating $X$ first, the partition function can be written
as a delta function along the line of  \cite{dj}: \bea Z & = &
\frac{1}{\mbox{Vol}(SO(N))} \left( \frac{\widetilde{\La}}{2 \pi
g_s}\right)^ {\frac{1}{2}N_f \widetilde{N}}
 \int  [dX] \prod_{j=1}^{N_f} [d q_j]
\de\left(X^{ij}-M^{ij}\right) e^{-\frac{1}{2g_s} \left(   \mbox{Tr} \; m X
+ \frac{1}{ \mu} \sum_{i,j=1}^{N_f} X^{ij} q_i \cdot q_j \right)}
\nonu \\
& = & \frac{1}{\mbox{Vol}(SO(N))} \left(\frac{\widetilde{\La}}{2
\pi g_s}\right)^ {\frac{1}{2}N_f \widetilde{N}} \int
\prod_{j=1}^{N_f} [d q_j]
 \de \left( \mu m + q_i \cdot q_j\right). \nonu
\eea
In the gauge theory side, this behavior of delta function
is equivalent to the $M^{ij}$
equations of motion led by $<q_i \cdot q_j>=-\mu m$.
The constrained matrix integral over $N_f$ flavors of length $\widetilde{N}$
can be obtained Wishart random matrices and the result \cite{dj,jn} for this
is
\bea
\int   \prod_{j=1}^{N_f} [d q_j]
\de \left( \mu m + q_i \cdot q_j\right) = c \times \left( \mbox{det}
\left(-\mu m\right)
\right)^{\left(\widetilde{N}-N_f-1\right)/2}
\nonu
\eea
where the coefficient $c$ behaves like as
\bea
e^{-\frac{1}{2} N_f \widetilde{N} \log \widetilde{\frac{N}{2}} }
\nonu
\eea
for large $\widetilde{N}$-limit.

As we did for electric case, the effective superpotential from the
log of partition function
can be expressed as
\bea
W & = & \frac{1}{2} \left(\widetilde{N}_c-2 \right)
\left( S - S \log \frac{S}{
{\widetilde{\La}}^3} \right)   -
\frac{1}{2} N_f
\left( S - S \log \frac{S}{ {\widetilde{\La}}^3} \right)
+
S N_f \log {\widetilde{\La}} - \frac{1}{2} S
\log \mbox{det} \left(-\mu m\right)
\nonu \\
& = &  \frac{1}{2} \left(\widetilde{N}_c-N_f-2 \right)
\left( S- S \log S + S \log \left( \frac{
\widetilde{\La}^{3\left(\widetilde{N}_c-2\right)-
N_f}}
{\mbox{det} \left(-\mu m\right)} \right)^{1/
\left(\widetilde{N}_c-N_f-2\right)} \right).
\label{W1}
\eea
Solving the F-flatness condition $\pa_S W=0$
one gets, with the phase factor $e^{2\pi i k/\left(
\widetilde{N}_c-N_f-2\right)},
k=1, \cdots, \left(\widetilde{N}_c-N_f-2\right)$
reflecting the $(\widetilde{N}_c-N_f-2)$ supersymmetric vacua,
\bea
S=\left( \frac{16 \widetilde{\La}^{3\left(\widetilde{N}_c-2\right)-
N_f}_{N_f-N_c+4,N_f}}{\mbox{det} \left(-\mu m\right)}
 \right)^{1/\left(\widetilde{N}_c-N_f-2\right)} \ep_{
\widetilde{N}_c-N_f-2}, \qquad
\widetilde{\La}^{3\left(\widetilde{N}_c-2\right)-
N_f}
= 16 \widetilde{\La}^{3\left(\widetilde{N}_c-2\right)-
N_f}_{N_f-N_c+4,N_f}.
\nonu
\eea

Then the exact superpotential by plugging this $S$ into (\ref{W1}) leads to
\bea
W & = & \frac{1}{2} \left(\widetilde{N}_c -N_f -2 \right) \ep_{
\widetilde{N}_c-N_f-2} \left(
\frac{ 16 \widetilde{\La}_{N_f-N_c+4,N_f}^
{3\left(\widetilde{N}_c-2\right)-N_f}}
{\mbox{det} \left(-\mu m\right) }\right)^{1/\left(\widetilde{N}_c-N_f-2\right)}
\nonu \\
& = & \frac{1}{2} \left(-N_c +2 \right)
\ep_{N_c-2}
\left(
\frac{ 16 \widetilde{\La}_{N_f-N_c+4,N_f}^{3\left(N_f-N_c+2\right)-N_f}}
{\left(-1\right)^{N_f} \mu^{N_f} \mbox{det} m }\right)^{-1/\left(N_c-2\right)}
\nonu \\
& = &
 \frac{1}{2} \left(N_c  -2 \right) \ep_{N_c-2} \left(
16  \La_{N_c,N_f}^{3\left(N_c-2\right)-N_f} \mbox{det}  m \right)^
{1/\left(N_c-2\right)}
\nonu \\
& = &
 \frac{1}{2} \left(N_c  -2 \right) \ep_{N_c-2}
\hat{\La}_{N_c-N_f,0}^3
\label{W2}
\eea
which is exactly the same as the one in (\ref{eW}).
Here we used the fact that
$\widetilde{N}_c = N_f -N_c +4$,
 $\mbox{det} \left(-\mu m\right) = \left(-1\right)^{N_f} \mu^{N_f}
\mbox{det} m$ and
$\ep_{
\widetilde{N}_c-N_f-2}=\ep_{-N_c+2}=\ep_{N_c-2}$.
The scale of the magnetic theory in the gauge theory side was
related to that of
electric theory by \cite{is}
\bea
2^8 \La_{N_c,N_f}^{3\left(N_c-2\right)-N_f}
\widetilde{\La}_{N_f-N_c+4,N_f}^{3\left(N_f-N_c+2\right)-N_f} =
\left(-1\right)^{N_f-N_c}
\mu^{N_f}
\nonu
\eea
where the normalization factor $1/2^8$ was chosen to get
the consistent low energy behavior under large mass deformation and along the
flat directions.
Note that the factor $(-1)^{N_c/(N_c-2)}=-1$ in (\ref{W2}) is cancelled exactly
by the overall $-1$ factor.
If not all of the matter fields are massive, the remaining low energy
magnetic theory is a magnetic $SO(N_f-N_c+4-M)$ gauge theory with $(N_f-M)$
flavors and the low energy superpotential. This low energy magnetic theory
is dual to the low energy $SO(N_c)$ gauge theory with  $(N_f-M)$ massless
quarks. One can arrive at the scale relation which connects the $SO(N_c)$
electric theory with $(N_f-M)$ flavors to the $SO(N_f-N_c+4-M)$ magnetic
theory with $(N_f-M)$ flavors.

\section{Discussions}

In this paper, we apply the matrix model for fundamental flavors without any
adjoint matter to check the Seiberg duality when we deform mass terms.
By explict matrix path integral in both electric and magnetic theories,
we have found the complete agreement with the field theory result.

One can consider and generalize degenerate mass deformation in which
some eigenvalues of the mass matrix $m$ are vanishing.
We expect to have the following superpotential, effectively
$SO(N_c)$ theory with $K$ flavors,
\bea
W =
\frac{1}{2} \left(N_c -K -2 \right) \ep_{N_c-K-2} \left(
\frac{16 \La_{N_c,K}^{3\left(N_c-2\right)-K} \mbox{det} m }{\mbox{det}
M }\right)^{1/\left(N_c-K-2\right)}
\nonu
\eea
where $M^{ij}(i,j= 1, 2, \cdots, K)$ are the meson from the massless flavors
and the index $i$ of mass matrix eigenvalues
$m_i$ runs  $i=K+1, K+2, \cdots, N_f$.
We have to add the terms from
the delta function constraint for massless flavor
part. For $K=0$, we can reproduce the result of section 2 and 3.

\newpage
\centerline{\bf Acknowledgments}

This research of CA was supported by
Korea Research Foundation Grant(KRF-2002-015-CS0006).
This research of SN was supported by
Korea Research Foundation Grant KRF-2001-041-D00049.

\end{document}